# FROM EXPECTATION TO EXPERIENCE: A BEFORE AND AFTER SURVEY OF PUBLIC OPINION ON AUTONOMOUS CARS IN SAUDI ARABIA

Mona Alfayez[a] ( 447205658@student.ksu.edu.sa ), and Dr. Ohoud Alharbi[b] ( omalharbi@ksu.edu.sa )

[a][b] *Software Engineering Department, College of Computer and Information Sciences, King Saud University, Riyadh, Kingdome of Saudi Arabia*

\* P.O. Box 22452, Riyadh 11495, Kingdom of Saudi Arabia

*Volume 1.0*
*12-22-2025*

# FROM EXPECTATION TO EXPERIENCE: A BEFORE AND AFTER SURVEY OF PUBLIC OPINION ON AUTONOMOUS CARS IN SAUDI ARABIA

**Abstract**

Autonomous vehicles (AVs) are emerging as a transformative innovation in transportation, offering potential benefits in safety, sustainability, and efficiency. Saudi Arabia's adoption of AVs aligns with Vision 2030, emphasizing smart mobility through initiatives such as the Riyadh Autonomous Metro and self-driving cars. This study explores Saudi citizens' perceptions of AVs before and after exposure to these technologies and examines whether demographic factors age, gender, education level, and driving habits affect acceptance. Using quantitative methods, the findings provide insights into the broader influences shaping AV adoption, highlighting the importance of trust, perceived safety, and convenience. These results can inform policymakers and industry stakeholders on strategies to facilitate successful integration of AVs into Saudi Arabia's transportation ecosystem.

**Introduction and Purpose:**

Autonomous Vehicles are one of the most significant enhancements in the transportation sector. Globally, autonomous vehicles are encouraged for their benefits in several aspects such as economics, safety, and sustainability. In recent years, governments and companies have been targeting autonomous vehicles as investments for their research and development as part of smart cities and intelligent transport systems (ITS). According to research, ITS is more likely to lower Carbon emissions, optimize logistics and modernize urban mobility.

Saudi Arabia moving toward autonomous vehicles is consistent with Vision 2030 emphasizing innovation, digital transformation and sustainable infrastructure. The launch of Riyadh Autonomous Metro, along with the gradual exposure of self driving cars are two main factors reflect the Kingdom's ambition to lead the way in smart mobility across the region, there are also other examples such as Al Haramain high speed railway, robotaxis and electric buses. Although technology and infrastructure are significant to this improvement in the transport sector, it also relies on Saudi Arabian's acceptance of these changes in their daily lives. Self driving cars have been trending among the residents of the Kingdom of Saudi Arabia. All these changes increase the experience of smart solutions, which could have the foremost impact on public opinion regarding autonomous and self driving cars. Along with these effects, there is a critical aspect of how culture and social interaction affect community judgment. In addition, the lack of research surveying the Saudi resident's opinion after the implementation of these major projects.

Studies done worldwide have proven that autonomous vehicles have a high positive impact on transportation cost reduction due to traffic management and fuel consumption. In addition to the economic consequences, autonomous vehicles influence the sustainability factors in a constructive way by lowering Carbon emissions. Furthermore, traffic safety is greatly enhanced if human interruptions are lowered due to smart mobility. For all those reasons and more, one of the strategic goals documented in Saudi Arabia 2030 Vision is to have at least 15% of public transportation vehicles operating autonomously and 25% of goods transport vehicles into autonomous vehicles by 2030. Although autonomous vehicles are being rapidly developed and introduced, their successful adoption depends not only on technical readiness but also on public acceptance.

A research (Aldakkhelallah *et al.*, 2023), conducted in Saudi Arabia, gives a comprehensive view on the acceptance and the expectation of the Saudi citizens regarding the autonomous and self drive cars. The research conducted at a time when mobility solutions were relatively new to the KSA public and the market. The findings of the research suggested growing awareness and generally positive attitudes toward adoption with education and gender significantly influenced adoption of AV technology.

Since 2023, the public experience of autonomous vehicles in Saudi Arabia has increased significantly, with self driving cars becoming increasingly popular due to their improved accessibility. The objective of this study is to comprehensively bridge the gap between the expectation and the reality of the Saudi community before and after key factor adjustments on the smart mobility sector have been launched and people being exposed to them. It aims to answer the following questions:

- How do Saudi citizens perceive autonomous cars after experiencing them?
- Do factors such as age, gender, education level, or weekly driving hours influence the acceptance of autonomous vehicles?

This research will provide insights for legislators and transport planners to design strategies for smoother autonomous vehicles adoption in Saudi Arabia.

## 2. Literature Review

### *2.1. Historical Context and Classification*

The development of autonomous vehicle (AV) technology has its roots in the early 20th century, with the "Linriccan Wonder" of the 1920s being the first radio controlled car (Beza *et al.*, 2022). This was followed by the showcase of electric cars powered by embedded circuits in 1939. A significant step toward modern autonomy occurred in 1980 with the advent of a robotic van by Mercedes Benz that utilized vision guided systems (Singh and Saini, 2021). This pioneering work marked the origin of foundational technologies like lane keep assist, lane departure warning, and adaptive cruise control that are present in modern vehicles today.

AVs are classified under Society of Automotive Engineers (SAE) into six levels (J. Wang *et al.*, 2021). These ranks are generally grouped into two major categories based on who performs the dynamic driving task primarily distinguishing between driver assisted and automated systems (Alqahtani, 2025). The first three levels 0 to 2 are classified as Driver Support, where at these levels the human driver must constantly supervise the system and perform the driving task, even when features like adaptive cruise control or combined steering and acceleration and braking are active. Conditional Automation is level 3, where the system handles all driving but requires humans to be ready to intervene. High Automation is level 4, where the vehicle operates without human intervention in a defined area, and level 5 is Full Automation, where the vehicle can drive itself under all conceivable road and environmental conditions.

The adoption and diffusion of any emerging technology are influenced by the degree of acceptance and trust demonstrated by the community and requires a great efforts by researchers, practitioners, and policymakers (Sedat Sonko *et al.*, 2024). This principle equally applies to AVs. A multitude of factors can significantly shape the rate at which AVs are adopted or resisted by the public. Individual psychological traits such as emotional and social dimensions influence the willingness to use AVs (Franziska Schandl, Peter Fischer, and Matthias F. C. Hudecek, 2023; Panagiotopoulos, Dimitrakopoulos and Keraite, 2024). Marketing and design efforts should be tailored to address the specific psychological needs and preferences of the diverse user profiles. AV systems should integrate adaptive, driver aligned behaviour models that enhance user experience without compromising safety (Ma and Zhang, 2021).

A research (Nacer Eddine Bezai *et al.*, 2021) performed a systematic review to identify the major obstacles hindering the full rollout of AVs, which are expected to revolutionize future cities. The research groups these challenges into two main areas. The first is focused on people and policy, covering issues like whether users will accept AVs, concerns about safety, and the urgent need for new laws and regulations. The second area focuses on technology, specifically highlighting barriers such as challenges with software and hardware, the necessity of advanced communication systems Vehicle to Everything (V2X) for cars to talk to each other and infrastructure, and the difficulty of ensuring highly accurate mapping and positioning. Another study (Campisi *et al.*, 2021) emphasizes on the need for public education, ethical considerations, and smart technologies strategies to increase AV adoption.

## 2.2. Saudi vs. Global Acceptance of Autonomous Vehicles

Few studies have been conducted to examine the adoption and acceptance of AVs among the population of Saudi Arabia. Two studies (Ibrahim Alsghan *et al.*, 2021), (Uneb Gazder and Eman Algherbal, 2025) on consumer acceptance of AVs in Saudi Arabia reported similar findings. Results are summarized as follows: research conducted among participants in Riyadh consistently shows a positive initial attitude towards technological advancements and a general concern for environmental issues. However, this willingness to adopt AVs is moderated by significant concerns over safety and the technology's performance under local weather conditions. Both studies identify trust in automakers and technological familiarity as key psychological determinants of acceptance.

Another research (Ibrahem Shatnawi *et al.*, 2025) assessed the Kingdom of Saudi Arabia's readiness for deploying AVs and Connected Autonomous Vehicles (CAVs) across five key areas: policy, technology, infrastructure, consumer acceptance, and social/environmental impact, based on a survey of local experts. The paper concludes that, despite active pilot projects and strong governmental support, accelerated effort in these lagging areas is necessary for successful, large scale, and safe AV-CAV deployment.

Another research (Aldakkhelallah *et al.*, 2023) investigated the public perception of AVs, using a survey conducted in Saudi Arabia to gather data on community awareness, attitudes, and readiness for the technology's introduction. The study used a mixed-method approach with 108 Saudi respondents, revealing that education strongly influences awareness of smart mobility and most prefer partial automation over full autonomy. 50% planned to buy an autonomous car within 10 years, while 40% expected 2030 to be the transition year. These findings indicate that the participants are largely receptive to new technologies and hold favourable attitudes toward the transition to AVs. However, the study revealed a wide variation in public opinion regarding the expected benefits, and that the perceived cost, convenience, and safety have a substantial impact on participants' opinions.

To determine whether Saudi citizens exhibit distinct reactions toward AVs compared to those in other countries, a comprehensive analysis of previous studies was conducted examining public responses in the United Kingdom, the United States, Australia, Japan, Europe, China, Pakistan, South Korea, and India. Public attitudes toward autonomous vehicles vary across countries, in the UK and US, and Australia public responses tend to foreground safety concerns more regularly than hopes for safety improvements (Matin and Dia, 2024; Tennant *et al.*, 2025). In Japan generally expressing favourable views, and Germany exhibiting negative perceptions (Taniguchi *et al.*, 2022). Remaining Europeans links acceptance to broader goals like environmental sustainability, safety, and ethical governance (Yang *et al.*, 2025). Chinese, Pakistanis, South Korea and Indians consumers' acceptance is driven by policy support, infrastructure readiness, and positive real-world applications, though concerns about safety and data security persist (Z. Wang *et al.*, 2021; Bhardwaj *et al.*, 2021; Maeng *et al.*, 2021).

The comparative analysis reveals that Saudi citizens exhibit a generally positive and receptive attitude toward AVs, mirroring the optimism seen in Japan and other markets like China, Pakistan, South Korea and India, but contrasting with the lukewarm reactions observed in the UK, US, Australia and EU.

## *2.3. Key Technical and Human Challenges*

A study (Ie Rei Park *et al.*, 2024) surveyed 2000 people, the findings revealed that acceptance is influenced by awareness, general reputation, perception, or prestige associated with the new technology, conditional value, and perceived risks. Research (Jack Stilgoe, 2021) shares similar findings, particularly in focusing on the non-technical, human centric factors that determine public acceptance of AVs.

A research (Shanzhi Chen *et al.*, 2024) systematically reviewed the critical technological domain of decision making and planning for AVs, specifically focusing on complex and challenging intersection environments. Intersections are highlighted as a major source of accidents and traffic congestion, making their safe and efficient negotiation a key requirement for full AV deployment. Cybersecurity challenges within AVs, specifically whether existing security measures are adequate for the technology's rapid deployment (Irmina Durlik *et al.*, 2024). A significant public concerns that act as barriers, specifically citing privacy including location tracking, and surveillance and security vulnerability to hackers (Girdhar *et al.*, 2023; Saeed *et al.*, 2023). In addition, reliability as an internal risk of Automated Driving System (ADS) and visibility via weather as an external risk influence a driver's trust in AV (Azevedo-Sa *et al.*, 2021).

The study by (Fayez Alanazi, 2023), addressed the critical need for a robust Intelligent Transportation System (ITS) infrastructure to support smart mobility solutions, which are vital for addressing traffic congestion, safety, and transportation management in the Kingdom of Saudi Arabia.

In this literature review, we took an overview of the history behind AV technology. In addition we looked to universal public acceptance of AVs and what it could be impacted by such as safety concerns, ethical considerations, legal liability, and regulatory frameworks (Othman, 2021). These factors contribute to a deeper understanding of how it influences the successful integration of AV and it gives us an idea of what could be an aspect affecting the opinions of Saudi citizens.

## 3. Study design/methodology/approach

The researcher designed this study to employ a quantitative approach using a survey to gain a comprehensive understanding of public opinion on autonomous vehicles in Saudi Arabia. The researcher distributed the survey in English and Arabic to reach wider range in Saudi Arabia. The questionnaire included 23 questions combined between demographic questions, 2 questions to measure responders previous knowledge of AVs and questions from 7 – 23 that was previously used in a related study by (Ibrahim Alsghan *et al.*, 2021) for a comparison between the AVs and the regular vehicles to measure Early Adoption Attitudes towards AVs, environmental concern, trust in AVs, Willing to Use AVs. The questions were rated on a Likert scale of 1 (strongly disagree) to 5 (strongly agree) except for the 7th question, which range from 1 (extremely uncomfortable) to 5 (extremely comfortable). The Likert scale results used to generate measurable data on public attitudes, trust levels, and perceived benefits of autonomous vehicles. The survey was distributed online. The study follows ethical research standards. Participants are informed about the purpose of the research. A non probability convenience sampling method was used for the survey, targeting participants from Riyadh, Saudi Arabia. Data collected from the survey was analysed using Microsoft Excel for quantitative components.

## 4. Findings

The survey was completed by 50 respondents. The demographic profile of the participants is summarized below in Table 1, providing context for the subsequent analysis.

| Demographic Factor | Category | Frequency (n) | Percentage (%) |
|---|---|---|---|
| **Age** | 35–44 | 17 | 34.0% |
| | 25–34 | 16 | 32.0% |
| | 45–54 | 6 | 12.0% |
| | 55–64 | 6 | 12.0% |
| | 18–24 | 3 | 6.0% |
| | 65+ | 2 | 4.0% |
| **Gender** | Female | 27 | 54.0% |
| | Male | 23 | 46.0% |
| **Level of Education** | Graduated from college | 35 | 70.0% |
| | Diploma | 6 | 12.0% |

|  | Master's degree | 6 | 12.0% |
|---|---|---|---|
|  | Ph.D. degree | 3 | 6.0% |
| **Weekly Driving Hours** | Between 6 and 10 hours | 13 | 26.0% |
|  | Less than 5 hours | 12 | 24.0% |
|  | I do not drive | 9 | 18.0% |
|  | Between 11 and 15 hours | 6 | 12.0% |
|  | Between 16 and 20 hours | 5 | 10.0% |
|  | More than 30 hours | 3 | 6.0% |
|  | Between 26 and 30 hours | 1 | 2.0% |
|  | Between 21 and 25 hours | 1 | 2.0% |

Table 1: Description of survey respondents

The sample is predominantly composed of individuals in the 25–44 age range (66.0%), with a slightly higher representation of females (54.0%). A significant majority of respondents hold a college degree (70.0%), indicating a highly educated sample. In terms of driving habits, the majority of respondents drive 10 hours or less per week (50.0%), with a notable portion (18.0%) reporting that they do not drive at all and rely on other transportation methods such as the Autonomous Metro.

| Topic | Q# | Question |
|---|---|---|

| **Early Adoption Attitudes** | 1 | How would you feel about driving on roads alongside autonomous (driverless) cars? |
|---|---|---|
| | 2 | Autonomous vehicle driving will be easier than manual driving |
| | 3 | I believe that within 30 years from now, automated driving system will be so advanced that is irresponsible to drive manually |
| **Environmental Concern** | 4 | I am willing to spend a bit more to buy a product that is more environmentally friendly |
| | 5 | I rarely worry about the effects of pollution on myself and my family |
| | 6 | I'm very concerned about current environmental pollution in Saudi Arabia and its impact on health |
| | 7 | I don't change my behaviour based solely on concern for the environment |
| **Trust** | 8 | I trust autonomous vehicles and I would like my family to use them |
| | 9 | Using autonomous vehicles will decrease my crash risk |
| | 10 | I will switch to manual driving from automated driving in case of poor weather |

|  | 11 | As a point of principle, humans should be in control of their vehicles at all times |
|---|---|---|
| **Willing to Use** | 12 | Autonomous vehicles will let me do other tasks, such as eating, watch a movie, be on a cell phone on my trip |
|  | 13 | I think Driving in congested areas is stressful |
|  | 14 | I find autonomous vehicles to be useful when I'm not feeling well |
|  | 15 | Using autonomous vehicles will be useful in meeting my driving needs |
|  | 16 | Using an autonomous vehicle would enable me to reach my destination safely |
|  | 17 | I find autonomous vehicles to be useful when I'm impaired |

Table 2: Survey questions related to AV acceptance

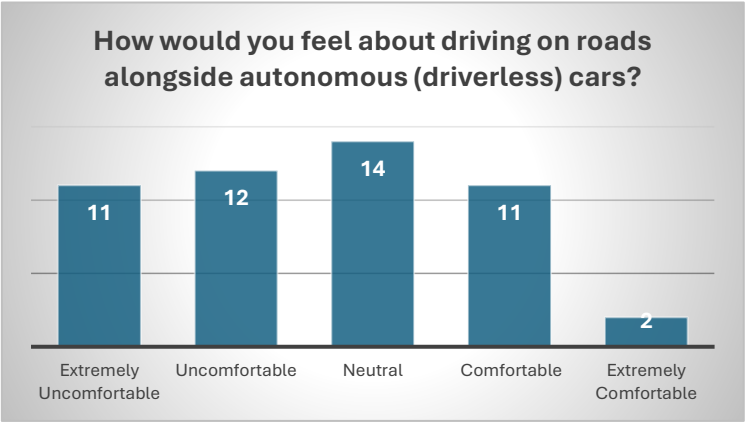

Figure 1: Distribution of responses on 1st question related to acceptance of AVs

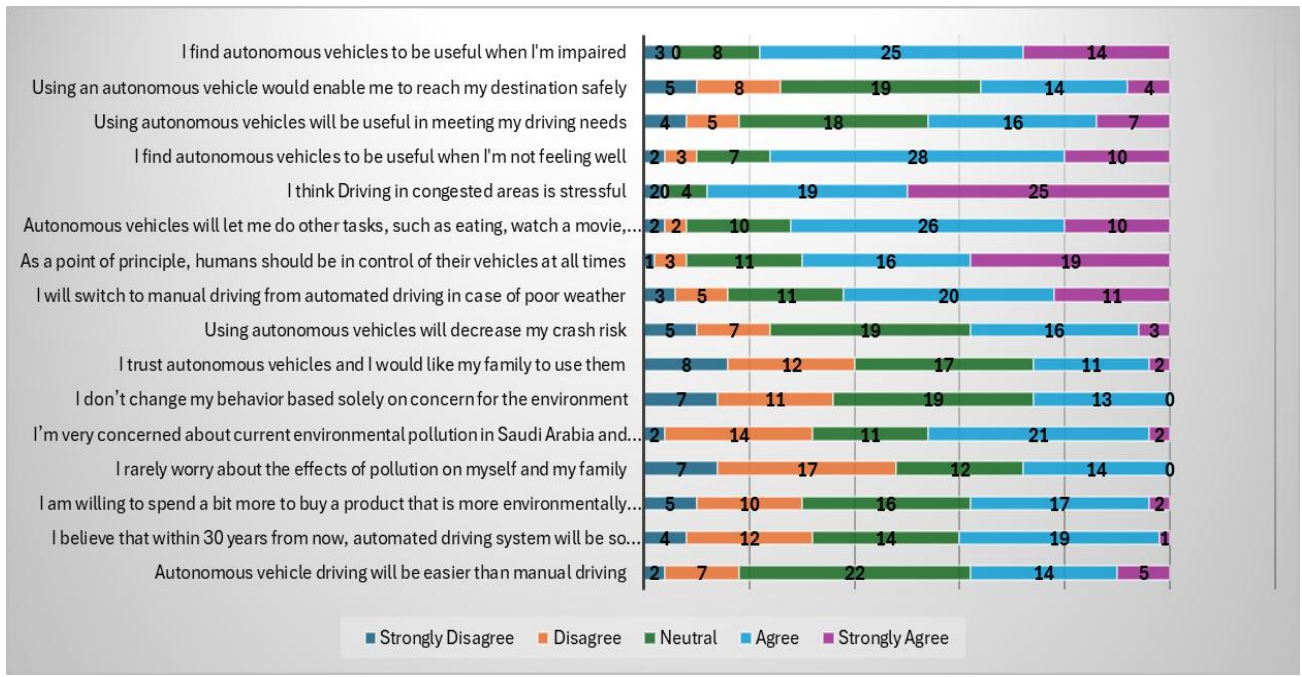

Figure 2: Distribution of responses on questions related to acceptance of AVs

The mean and standard deviation for each survey item, measured on a 5-point Likert scale (1 = Strongly Disagree/Extremely Uncomfortable, 5 = Strongly Agree/Extremely Comfortable), are presented in Table 3.

| **Question** | **Mean** | **SD** | **1** | **2** | **3** | **4** | **5** |
|---|---|---|---|---|---|---|---|
| How would you feel about driving on roads alongside autonomous (driverless) cars? | 2.62 | 1.18 | 11 | 12 | 14 | 11 | 2 |
| Autonomous vehicle driving will be easier than manual driving | 3.26 | 0.97 | 2 | 7 | 22 | 14 | 5 |

| Statement | Mean | SD | 1 | 2 | 3 | 4 | 5 |
|---|---|---|---|---|---|---|---|
| I believe that within 30 years from now, automated driving system will be so advanced that is irresponsible to drive manually | 3.02 | 1.02 | 4 | 12 | 14 | 19 | 1 |
| I am willing to spend a bit more to buy a product that is more environmentally friendly | 3.02 | 1.06 | 5 | 10 | 16 | 17 | 2 |
| I rarely worry about the effects of pollution on myself and my family | 2.66 | 1.04 | 7 | 17 | 12 | 14 | 0 |
| I'm very concerned about current environmental pollution in Saudi Arabia and its impact on health | 3.14 | 1.01 | 2 | 14 | 11 | 21 | 2 |
| I don't change my behaviour based solely on concern for the environment | 2.76 | 1.001 | 7 | 11 | 19 | 13 | 0 |
| I trust autonomous vehicles, and I would like my family to use them | 2.74 | 1.10 | 8 | 12 | 17 | 11 | 2 |
| Using autonomous vehicles will decrease my crash risk | 3.1 | 1.06 | 5 | 7 | 19 | 16 | 3 |
| I will switch to manual driving from automated driving in case of poor weather | 3.62 | 1.12 | 3 | 5 | 11 | 20 | 11 |
| As a point of principle, humans should be in control of their vehicles at all times | 3.98 | 1.02 | 1 | 3 | 11 | 16 | 19 |
| Autonomous vehicles will let me do other tasks, such as eating, watch a movie, be on a cell phone on my trip | 3.8 | 0.95 | 2 | 2 | 10 | 26 | 10 |
| I think Driving in congested areas is stressful | 4.3 | 0.93 | 2 | 0 | 4 | 19 | 25 |
| I find autonomous vehicles to be useful when I'm not feeling well | 3.82 | 0.96 | 2 | 3 | 7 | 28 | 10 |
| Using autonomous vehicles will be useful in meeting my driving needs | 3.34 | 1.09 | 4 | 5 | 18 | 16 | 7 |
| Using an autonomous vehicle would enable me to reach my destination safely | 3.08 | 1.08 | 5 | 8 | 19 | 14 | 4 |
| I find autonomous vehicles to be useful when I'm impaired | 3.94 | 0.99 | 3 | 0 | 8 | 25 | 14 |

Table 3: Descriptive Statistics for Survey Items

## 4.1. Early AVs Adoption Attitudes compared with Traditional Vehicles

Questions 1–3 focused on respondents' opinions regarding autonomous vehicles (AVs) compared to traditional vehicles to measure early adoption attitudes. The results indicate that most participants agreed that AVs would make driving easier than manual driving (Question 2, Mean = 3.26). There was also moderate agreement with the statement that automated driving systems will become so advanced that manual driving will be considered irresponsible (Question 3, Mean = 3.02). However, the lowest rating was observed for Question 1 (Mean = 2.62), suggesting discomfort with driving alongside AVs. This implies that while respondents recognize the potential benefits of AVs, concerns remain about sharing roads with them.

### 4.2. Consideration of Environmental Concerns

Questions 4–7 assessed environmental attitudes related to AV adoption. Respondents showed moderate agreement with paying more for environmentally friendly products (Question 4, Mean = 3.02) and concern about pollution in Saudi Arabia (Question 6, Mean = 3.14). Conversely, statements reflecting low environmental concern, such as rarely worrying about pollution (Question 5, Mean = 2.66) and not changing behaviour based on environmental issues (Question 7, Mean = 2.76), received lower ratings. These findings suggest that environmental awareness slightly influences AV acceptance.

### 4.3. Trust in Use of AV

Questions 8–11 explored trust in AVs. Ratings were mixed: respondents moderately agreed that AVs reduces crash risk (Question 9, Mean = 3.1) but expressed strong agreement that humans should maintain control at all times (Question 11, Mean = 3.98). This indicates that while trust in AV technology exists, there is a strong preference for human oversight, especially in adverse conditions.

### 4.4. Willingness to Use AV

Questions 12–17 measured willingness to use AVs in various scenarios. All items scored high, with the strongest agreement for using AVs in congested areas (Question 13, Mean = 4.3) and when not feeling well (Question 14, Mean = 3.82). Respondents also agreed that AVs would allow multitasking during trips (Question 12, Mean = 3.8) and provide safe travel when impaired (Question 17, Mean = 3.94). These results highlight that practical benefits, such as convenience and safety, significantly drive willingness to adopt AVs.

*4.5. Correlation Analysis*

The Pearson correlation matrix shows the relationship between the construct Willingness to Use AVs and the other factors: Early Adoption Attitudes, Environmental Concern and Trust.

The correlation analysis presented in Table 4 highlights several important relationships. The strongest positive association is observed between Early Adoption Attitudes and Willingness to Use AVs (r = 0.65), indicating that individuals who are enthusiastic about adopting new technologies are significantly more inclined to accept autonomous vehicles. Similarly, Trust and Willingness to Use AVs exhibit a strong positive correlation (r = 0.61), suggesting that confidence in AV technology plays a critical role in shaping adoption intent. In contrast, Environmental Concern shows a near-zero and slightly negative correlation with Willingness to Use AVs (r = -0.04), confirming that environmental attitudes, as measured in this study, do not substantially influence AV acceptance. These findings underscore that psychological factors such as trust and openness to innovation are far more influential than environmental considerations in determining willingness to adopt AVs.

|  | Willingness To Use Avs Correlation (r) |
|---|---|
| **Early Adoption Attitudes** | 0.65 |
| **Environmental Concern** | -0.04 |
| **Trust** | 0.61 |

Table 4: Pearson correlation coefficients between the construct Willingness to Use AVs and the other factors

As illustrated in Figure 3, the researcher employed a box plot to compare the distribution of ratings across four key factors influencing AV acceptance: willingness to use, trust, environmental concern, and early adoption attitudes. Among these, willingness to use exhibited the highest median score, indicating a generally positive inclination toward adopting AVs. However, the wide interquartile range and presence of outliers suggest substantial variability in participants' responses, reflecting a lack of consensus. Trust in AVs demonstrated a comparatively lower median rating, highlighting persistent concerns regarding reliability and safety. In contrast, environmental concern and early adoption attitudes displayed similar median values with relatively narrower spreads, indicating more consistent perceptions among respondents for these factors.

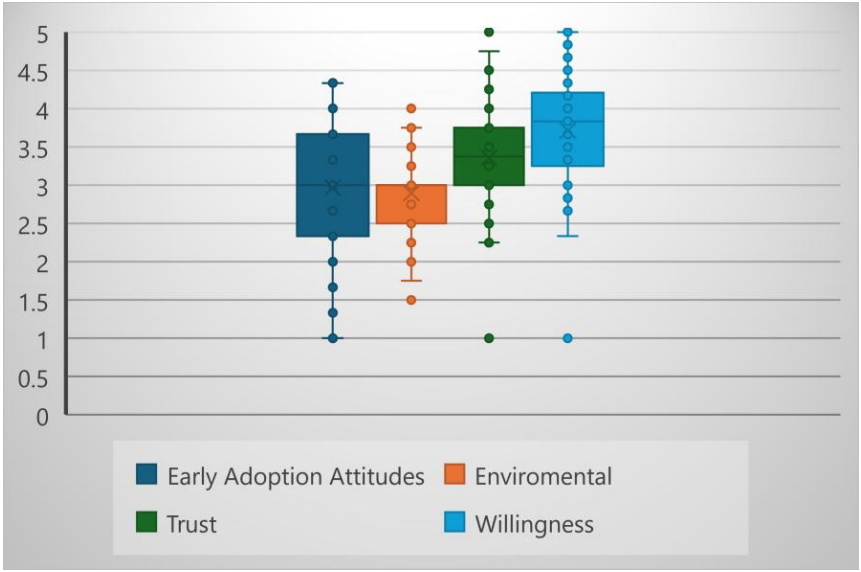

Figure 3: Box plot for comparison of ratings for different aspects of AVs acceptance

The researcher conducted a between subject design ANOVA test to examine whether age groups influence attitudes toward autonomous vehicle (AV) adoption across four constructs: Early Adoption Attitudes, Environmental, Trust, and Willingness to Use AVs. The results in Table 5 revealed no statistically significant differences among age groups for any construct, with p-values of 0.45 for Early Adoption Attitudes, 0.98 for Environmental, 0.72 for Trust, and 0.28 for Willingness. All values exceeded the 0.05 significance threshold, indicating that age does not play a significant role in shaping perceptions or acceptance of AVs within the sampled population.

| Construct | P-value |
| --- | --- |
| **Early Adoption Attitudes** | 0.45 |
| **Environmental** | 0.98 |
| **Trust** | 0.72 |
| **Willingness to use AVs** | 0.28 |

Table 5: ANOVA test for Age

The Author done a T-test to find if gender has any effect on any of the four factors (Early Adoption Attitudes, Environmental, Trust and Willingness to use AVs) and the results are in Table 6 below. The independent samples t-tests conducted to examine gender differences across the four constructs Early Adoption Attitudes, Environmental Concern, Trust, and Willingness to Use AVs revealed no statistically significant differences at the conventional α = 0.05 level. The p-values for all constructs exceeded 0.05, indicating that male and female respondents did not differ meaningfully in their ratings. Notably, the Trust construct approached significance (p ≈ 0.09), suggesting a potential trend toward higher trust among male participants compared to females. However, this trend does not meet the threshold for statistical significance and should be interpreted with caution. Overall, these findings imply that gender does not exert a substantial influence on attitudes toward AVs acceptance within this sample.

| Construct | P-value |
|---|---|
| **Early Adoption Attitudes** | 0.49 |
| **Environmental** | 0.25 |
| **Trust** | 0.09 |
| **Willingness to use AVs** | 0.35 |

Table 6: T-test for gender

As shown in Figure 4 below, Males reported slightly higher scores for all factors, with the largest difference observed in Willingness to Use (Male = 3.83; Female = 3.61) and Trust (Male = 3.53; Female = 3.21). Similarly, for Early Adoption Attitudes and Environmental Concern, male ratings (3.06 and 2.99, respectively) were marginally higher than those of females (2.89 and 2.81). These differences, however, were not statistically significant based on t-test results, suggesting that gender does not substantially influence attitudes toward autonomous vehicle acceptance in this sample. The small variations may indicate a trend toward greater confidence and willingness among male respondents, but this requires further investigation.

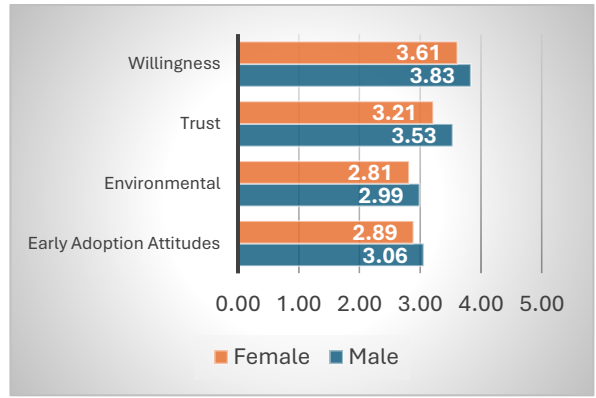

Figure 4: Comparison of Male and Female ratings

The ANOVA results for the four constructs in Table 7 indicate that there are no statistically significant differences among the groups being compared. Since all p-values are greater than the conventional threshold of 0.05, we fail to reject the null hypothesis for each construct. This suggests that the mean scores (Figure 5) across the different education levels (or other grouping factors used in the analysis) are relatively similar for all four constructs. In other words, education level does not appear to influence perceptions related to Early Adoption Attitudes, environmental considerations, trust, or willingness to adopt autonomous vehicles. These findings imply that attitudes toward these constructs are consistent across groups, and other demographic or contextual factors may play a more significant role in shaping these views.

| Construct | P-value |
|---|---|
| **Early Adoption Attitudes** | 0.44 |
| **Environmental** | 0.99 |
| **Trust** | 0.99 |
| **Willingness to use AVs** | 0.99 |

Table 7: ANOVA test for education level

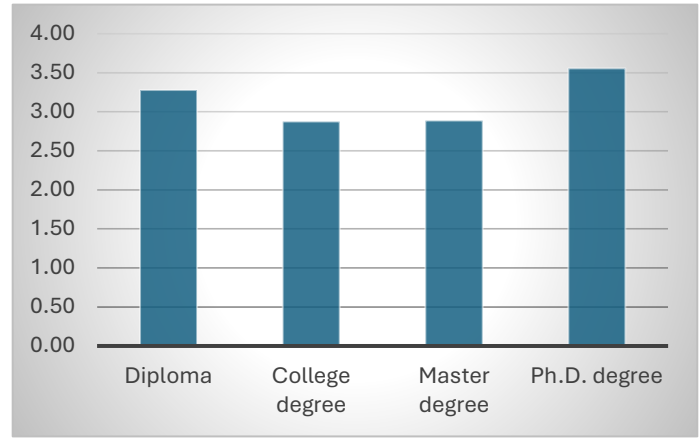

Figure 5: Mean scores of sample ratings grouped on education level

The researcher conducted another ANOVA test to find the significance of weekly driving hours of acceptance of AVs. Results are reported in Table 8 below. All p-values are greater than 0.05, which means there is no statistically significant difference in the mean scores of these constructs across different weekly driving hour groups. In other words, how many hours participants drive per week does not significantly affect their acceptance of autonomous vehicles in terms of Early Adoption Attitudes, Environmental considerations, Trust, or Willingness to use AVs. This suggests that driving experience, measured by weekly hours, may not be a key determinant of attitudes toward AVs adoption.

| Construct | P-value |
|---|---|
| **Early Adoption Attitudes** | 0.95 |
| **Environmental** | 0.31 |
| **Trust** | 0.94 |

| **Willingness to use AVs** | 0.98 |
|---|---|

Table 8: ANOVA test for weekly driving hours

*4.6. Comparison with Previous Studies*

To answer the first research question, comparison between a similar study on factors influencing autonomous vehicle (AV) acceptance was conducted in Riyadh, Saudi Arabia by (Ibrahim Alsghan *et al.,* 2021). Their research involved a large sample of 500 participants and examined constructs such as comparison which is named early adoption attitudes in this research, environmental concern, trust, tech-savviness, and willingness to use AVs. The study found that willingness to use AVs scored the highest among all factors, while trust in AVs was relatively low, indicating persistent concerns about safety and reliability. Furthermore, their parametric and predictive analyses revealed that age, trust, and being tech-savvy were significant determinants of willingness to adopt AVs, with younger participants showing greater acceptance. Gender differences were also observed, with women favouring AVs more than men, and education level having no significant effect on AVs acceptance.

In contrast, this study, also conducted in Saudi Arabia and after the launch of the AVs in Riyadh such as Riyadh Metro, Autonomous taxis and higher exposure of AVs overall, produced somewhat different insights. Our ANOVA results showed no statistically significant differences across demographic factors such as age, gender, education level, and weekly driving hours for any of the four constructs: Early Adoption Attitudes, environmental concern, trust, and willingness to use AVs (all p-values > 0.05). This suggests that, unlike (Ibrahim Alsghan *et al.*, 2021) findings, demographic characteristics in our sample do not appear to influence AVs acceptance. Both studies, however, agree on the importance of practical benefits and trust related concerns as underlying themes in AVs adoption.

The divergence in findings may be attributed to differences in sample size, demographic composition, and timing of the studies. The research (Ibrahim Alsghan *et al.*, 2021) highlighted the role of tech-savviness and age, whereas our results indicate a more uniform attitude toward AVs across demographic groups. This could reflect evolving perceptions as AV technology becomes more familiar or a sample specific trend. Future research should explore psychological and contextual factors such as perceived safety, cost, and infrastructure readiness that may better explain AVs acceptance in Saudi Arabia. Additionally, longitudinal studies could help determine whether demographic influences diminish over time as exposure to AV technology increases.

This contrasts with global research, where demographic and cultural differences often play a notable role. For example, studies in the UK, US, and Australia report strong safety concerns and scepticism toward AVs, while European acceptance is linked to environmental sustainability and ethical governance. Conversely, countries such as Japan, China, India, Pakistan, and South Korea exhibit higher optimism, driven by policy support and perceived convenience, despite lingering concerns about data security. Compared to these global trends, Saudi respondents share similarities with Asian markets in their generally positive outlook toward AVs but differ from Western nations where trust and safety concerns dominate. These comparisons highlight that while global attitudes vary widely, Saudi Arabia's acceptance appears less influenced by demographics and more aligned with broader technological optimism seen in emerging markets.

## *4.7. Effect of Factors on Acceptance of AVs*

The findings from this research provide valuable insights into the factors affecting the acceptance of AVs among the respondents. These findings contribute to answer the second research question.

The descriptive statistics, based on raw scores, clearly illustrate a dual perspective. On one hand, there is a strong appreciation for the practical utility and convenience of AVs, particularly in mitigating the stress of congested driving and providing mobility when the driver is impaired or unwell. This high agreement with utility focused statements suggests that respondents are receptive to the functional benefits of the technology.

On the other hand, the high raw mean scores for "humans should be in control of their vehicles at all times" and the intention to "switch to manual driving in case of poor weather" reveal a profound psychological barrier rooted in a preference for human agency and a fundamental lack of trust in the technology's reliability in non ideal conditions. This indicates that while respondents see the potential benefits of AVs, they remain highly sceptical of their safety and reliability in critical driving scenarios.

The correlation analyses reveal that Early Adoption Attitudes and Willingness to Use AVs are strongly related, underscoring the importance of openness to innovation in shaping acceptance. Trust also shows a robust positive association with willingness, highlighting confidence in AV technology as a critical driver of adoption. In contrast, Environmental Concern demonstrates a negligible and slightly negative correlation, suggesting that ecological attitudes, as measured here, are not meaningful predictors of AV acceptance. These results emphasize that psychological constructs such as trust and enthusiasm for new technologies are far more influential than environmental considerations in determining willingness to adopt autonomous vehicles.

Finally, Contrary to expectations and previous research, demographic variables such as age, gender, education level, and weekly driving hours did not show any statistically significant effect on attitudes toward AVs adoption across the four constructs examined: Early Adoption Attitudes, environmental concern, trust, and willingness to use AVs. All ANOVA and t-test results yielded p-values greater than 0.05, indicating that these demographic characteristics do not substantially shape perceptions of AVs within the sampled population.

The absence of significant differences across age groups suggests that demographic factors such as age may not be critical determinants of AVs acceptance. This finding implies that attitudes toward AVs are relatively uniform across generations, possibly due to shared concerns about safety, reliability, and technological readiness. For policymakers and industry stakeholders, this means that promotional strategies and educational campaigns do not need to be age specific but should instead focus on universal factors such as improving trust in technology, demonstrating safety benefits, and highlighting environmental advantages.

## 5. Research Implications

The uniformity of attitudes across demographic groups has practical implications for policymakers and industry stakeholders. Marketing campaigns and awareness programs do not need to be tailored to specific age or gender segments; rather, they should focus on building trust, highlighting safety benefits, and demonstrating practical advantages such as convenience and reduced stress in congested traffic. Additionally, strategies should address cybersecurity and data privacy concerns, which previous studies have linked to trust issues in AVs adoption.

Future studies should explore psychological factors (e.g., perceived risk, technology anxiety), economic considerations (e.g., cost and willingness to pay), and contextual influences (e.g., infrastructure readiness, cultural attitudes) that may better explain variations in AVs acceptance. Incorporating qualitative methods, such as interviews or focus groups, could provide deeper insights into the reasons behind uniform demographic responses and uncover latent concerns or motivations.

## 6. Originality/value

This study examined the factors influencing the acceptance of autonomous vehicles (AVs) in Saudi Arabia, focusing on four key constructs: Early Adoption Attitudes, environmental concern, trust, and willingness to use AVs. Using quantitative methods, the analysis revealed that demographic variables age, gender, education level, and weekly driving hours do not significantly affect attitudes toward AVs adoption. All p-values exceeded the conventional threshold of 0.05, indicating uniform perceptions across these groups.

These findings suggest that demographic characteristics may not be critical determinants of AVs acceptance in the Saudi context. Instead, attitudes toward AVs appear to be shaped by broader factors such as trust in technology, perceived safety, and convenience rather than personal attributes. This insight has practical implications for policymakers and industry stakeholders: awareness campaigns and marketing strategies should prioritize building trust and demonstrating safety and usability rather than targeting specific demographic segments.

Ultimately, this study contributes to the growing body of literature on AVs acceptance by highlighting the minimal role of demographic factors and emphasizing the need to focus on psychological and contextual influences. These findings can inform strategies to facilitate the successful integration of AVs into Saudi Arabia's transportation ecosystem.

**Acknowledgment:**

The author would like to acknowledge the support of King Saud University (KSU).

**Declaration of Interest:**

The author declares no conflict of interest.

**Article Word Count:**

7555 word

**References:**

Aldakkhelallah, A., Alamri, A.S., Georgiou, S., Simic, M.,(2023) 'Public Perception of the Introduction of Autonomous Vehicles', *World Electric Vehicle Journal*, Vol. 14 No. 12, p. 345. Available at: https://doi.org/10.3390/wevj14120345.

Alqahtani, T. (2025) 'Recent Trends in the Public Acceptance of Autonomous Vehicles: A Review', *Vehicles*, Vol. 7 No. 2, p. 45. Available at: https://doi.org/10.3390/vehicles7020045.

Azevedo-Sa, H., Zhao, H., Esterwood, C., Yang, X.J., Tilbury, D.M., Robert, L.P., (2021) 'How internal and external risks affect the relationships between trust and driver behavior in automated driving systems', *Transportation Research Part C: Emerging Technologies*, Vol. 123, p. 102973. Available at: https://doi.org/10.1016/j.trc.2021.102973.

Beza, A.D., Koutra, S., John, F., Thomas, D., (2022) 'The Emergence of Automated Vehicles and their Potential Implications for Urban Mobility: A Review of the Literature and Synthesis of Use Cases'. ENGINEERING. Available at: https://doi.org/10.20944/preprints202207.0448.v1.

Bhardwaj, A., Kumar, Y. and Hasteer, N. (2021) 'A Hybrid Model to Investigate Perception Towards Adoption of Autonomous Vehicles in India', in *2021 IEEE 4th International Conference on Computing, Power and Communication Technologies (GUCON). 2021 IEEE 4th International Conference on Computing, Power and Communication Technologies (GUCON)*, Kuala Lumpur, Malaysia: IEEE, pp. 1–5. Available at: https://doi.org/10.1109/GUCON50781.2021.9573588.

Campisi, T., Severino, A., Al-Rashid, M.A., Pau, G., (2021) 'The Development of the Smart Cities in the Connected and Autonomous Vehicles (CAVs) Era: From Mobility Patterns to Scaling in Cities', *Infrastructures*, Vol. 6 No. 7, p. 100. Available at: https://doi.org/10.3390/infrastructures6070100.

Fayez Alanazi (2023) 'Development of Smart Mobility Infrastructure in Saudi Arabia: A Benchmarking Approach'. Available at: https://doi.org/10.3390/su15043158.

Franziska Schandl, Peter Fischer, and Matthias F. C. Hudecek (2023) 'Predicting acceptance of autonomous shuttle buses by personality profiles: a latent profile analysis'. Available at: https://doi.org/10.1007/s11116-023-10447-4.

Girdhar, M., Hong, J. and Moore, J. (2023) 'Cybersecurity of Autonomous Vehicles: A Systematic Literature Review of Adversarial Attacks and Defense Models', *IEEE Open Journal of Vehicular Technology*, Vol. 4, pp. 417–437. Available at: https://doi.org/10.1109/OJVT.2023.3265363.

Ibrahem Shatnawi, Juan Nicolas Gonzalez, and Neda Masoud (2025) 'Toward AV-CAV deployment in the Kingdom of Saudi Arabia: A readiness assessment based on expert feedback'. Available at: https://doi.org/10.1016/j.rtbm.2025.101378.

Alsghan, I., Gazder, U., Assi, K., Hakem, G.H., Sulail, M.A., Alsuhaibani, O.A., (2021) 'The Determinants of Consumer Acceptance of Autonomous Vehicles: A Case Study in Riyadh, Saudi Arabia'. Available at: https://doi.org/10.1080/10447318.2021.2002046.

Ie Rei Park, Seoyong KimID, and Jungwook Moon (2024) 'Why do people resist AI-based autonomous cars?: Analyzing the impact of the risk perception paradigm and conditional value on public acceptance of autonomous vehicles'. Available at: https://doi.org/10.1371/journal.pone.0313143.

Irmina Durlik, Tymoteusz Miller, Ewelina Kostecka, Zenon Zwierzewicz, Adrianna Łobodzí nska, (2024) 'Cybersecurity in Autonomous Vehicles—AreWe Ready for the Challenge?' Available at: https://doi.org/10.3390/%2520electronics13132654.

Jack Stilgoe (2021) 'How can we know a self-driving car is safe?' Available at: https://doi.org/10.1007/s10676-021-09602-1.

Ma, Z. and Zhang, Y. (2021) 'Drivers trust, acceptance, and takeover behaviors in fully automated vehicles: Effects of automated driving styles and driver's driving styles', *Accident Analysis & Prevention*, Vol. 159, p. 106238. Available at: https://doi.org/10.1016/j.aap.2021.106238.

Maeng, K., Jeon, S.R., Park, T., Cho, Y., (2021) 'Network effects of connected and autonomous vehicles in South Korea: A consumer preference approach', *Research in Transportation Economics*, Vol. 90, p. 100998. Available at: https://doi.org/10.1016/j.retrec.2020.100998.

Matin, A. and Dia, H. (2024) 'Public Perception of Connected and Automated Vehicles: Benefits, Concerns, and Barriers from an Australian Perspective', *Journal of Intelligent and Connected Vehicles*, Vol. 7 No. 2, pp. 108–128. Available at: https://doi.org/10.26599/JICV.2023.9210028.

Bezai, N.E., Medjdoub, B., Al-Habaibeh, A., Chalal, M.L., Fadli, F., (2021) 'Future cities and autonomous vehicles: analysis of the barriers to full adoption'. Available at: https://doi.org/10.1016/j.enbenv.2020.05.002.

Othman, K. (2021) 'Public acceptance and perception of autonomous vehicles: a comprehensive review', *AI and Ethics*, Vol. 1 No. 3, pp. 355–387. Available at: https://doi.org/10.1007/s43681-021-00041-8.

Panagiotopoulos, I.E., Dimitrakopoulos, G.J. and Keraite, G. (2024) 'On Modelling and Investigating User Acceptance of Highly Automated Passenger Vehicles', *IEEE Open Journal of Intelligent Transportation Systems*, Vol. 5, pp. 70–84. Available at: https://doi.org/10.1109/OJITS.2023.3346477.

Saeed, Z., Masood, M. and Khan, M.U. (2023) 'A Review: Cybersecurity Challenges and their Solutions in Connected and Autonomous Vehicles (CAVs)', *JAREE (Journal on Advanced Research in Electrical Engineering)*, Vol. 7 No. 1. Available at: https://doi.org/10.12962/jaree.v7i1.322.

Sedat Sonko, Emmanuel Augustine Etukudoh, Kenneth Ifeanyi Ibekwe, Valentine Ikenna Ilojianya, Cosmas Dominic Daudu, (2024) 'A comprehensive review of embedded systems in autonomous vehicles: Trends, challenges, and future directions', *World Journal of Advanced Research and Reviews*, Vol. 21 No. 1, pp. 2009–2020. Available at: https://doi.org/10.30574/wjarr.2024.21.1.0258.

Chen, S., Hu, X., Zhao, J., Wang, R., Qiao, M., (2024) 'A Review of Decision-Making and Planning for Autonomous Vehicles in Intersection Environments'. Available at: https://doi.org/10.3390/wevj15030099.

Singh, S. and Saini, B.S. (2021) 'Autonomous cars: Recent developments, challenges, and possible solutions', *IOP Conference Series: Materials Science and Engineering*, Vol. 1022 No. 1, p. 012028. Available at: https://doi.org/10.1088/1757-899X/1022/1/012028.

Taniguchi, A., Enoch, M., Theofilatos, A., Ieromonachou, P., (2022) 'Understanding acceptance of autonomous vehicles in Japan, UK, and Germany', *Urban, Planning and Transport Research*, Vol. 10 No. 1, pp. 514–535. Available at: https://doi.org/10.1080/21650020.2022.2135590.

Tennant, C., Stares, S., Howard, S., (2025) 'Public anticipations of self-driving vehicles in the UK and US', *Mobilities*, Vol. 20 No. 2, pp. 292–309. Available at: https://doi.org/10.1080/17450101.2024.2325386.

Uneb Gazder and Eman Algherbal (2025) 'Determining Factors Affecting Acceptance of Autonomous Vehicles using Statistical and Machine Learning Models'. Available at: https://doi.org/10.48088/ejg.u.gaz.16.2.225.240.

Wang, J., Huang, H., Li, K., Li, J., (2021) 'Towards the Unified Principles for Level 5 Autonomous Vehicles', *Engineering*, Vol. 7 No. 9, pp. 1313–1325. Available at: https://doi.org/10.1016/j.eng.2020.10.018.

Wang, Z., Safdar, M., Zhong, S., Liu, J., Xiao, F., (2021) 'Public Preferences of Shared Autonomous Vehicles in Developing Countries: A Cross-National Study of Pakistan and China', *Journal of Advanced Transportation*. Edited by R.-Y. Guo, 2021, pp. 1–19. Available at: https://doi.org/10.1155/2021/5141798.

Yang, Y., Peng, L. and Wan, D. (2025) 'A Comparative Study on the Acceptance of Autonomous Driving Technology by China and Europe: A Transnational Empirical Analysis Based on the Technology Acceptance Model'. Engineering. Available at: https://doi.org/10.20944/preprints202508.1473.v1.

# Appendix A: Survey Questions

| Topic | Question Number | Question |
|---|---|---|
| **Demographics** | 1 | Age<br><br>18–24<br><br>25–34<br><br>35–44<br><br>45–54<br><br>55–64<br><br>65+ |
| | 2 | Gender<br><br>Female<br><br>Male |
| | 3 | Level of education<br><br>Did not attend school<br><br>Graduated from high school<br><br>Diploma<br><br>Graduated from college<br><br>Master degree<br><br>Ph.D. degree |

|  | 4 | Hours driving weekly |
|  |  | I do not drive but I use alternative options (Autonomous metro, AV, ….) |
|  |  | Less than 5 hours |
|  |  | Between 6 and 10 hours |
|  |  | Between 11 and 15 hours |
|  |  | Between 16 and 20 hours |
|  |  | Between 21 and 25 hours |
|  |  | Between 26 and 30 hours |
|  |  | More than 30 hours |
| **Knowledge of AVs** | 5 | How familiar are you with the concept of autonomous (driverless) vehicles? |
|  |  | Not familiar at all |
|  |  | Slightly familiar |
|  |  | Moderately familiar |
|  |  | Very familiar |
|  |  | Extremely familiar |
|  | 6 | Which of the following best describes your understanding of autonomous vehicle technology? |
|  |  | I know nothing about how it works |
|  |  | I know basic information (e.g., it drives without human input) |
|  |  | I understand some technical aspects (e.g., sensors, AI, safety systems) |

| | | I have in-depth knowledge of autonomous vehicle systems |
|---|---|---|
| **Early Adoption Attitudes** | 7 | How would you feel about driving on roads alongside autonomous (driverless) cars? range from 1 (extremely uncomfortable) to 5 (extremely comfortable) |
| | 8 | Autonomous vehicle driving will be easier than manual driving scale of 1 (strongly disagree) to 5 (strongly agree) |
| | 9 | I believe that within 30 years from now, automated driving system will be so advanced that is irresponsible to drive manually scale of 1 (strongly disagree) to 5 (strongly agree) |
| **Environmental Concern** | 10 | I am willing to spend a bit more to buy a product that is more environmentally friendly scale of 1 (strongly disagree) to 5 (strongly agree) |
| | 11 | I rarely worry about the effects of pollution on myself and my family scale of 1 (strongly disagree) to 5 (strongly agree) |
| | 12 | I'm very concerned about current environmental pollution in Saudi Arabia and its impact on health scale of 1 (strongly disagree) to 5 (strongly agree) |
| | 13 | I don't change my behavior based solely on concern for the environment scale of 1 (strongly disagree) to 5 (strongly agree) |
| **Trust** | 14 | I trust autonomous vehicles and I would like my family to use them scale of 1 (strongly disagree) to 5 (strongly agree) |
| | 15 | Using autonomous vehicles will decrease my crash risk |
|  |  | scale of 1 (strongly disagree) to 5 (strongly agree) |
|  | 16 | I will switch to manual driving from automated driving in case of poor weather<br><br>scale of 1 (strongly disagree) to 5 (strongly agree) |
|  | 17 | As a point of principle, humans should be in control of their vehicles at all times<br><br>scale of 1 (strongly disagree) to 5 (strongly agree) |
| **Willing to Use** | 18 | Autonomous vehicles will let me do other tasks, such as eating, watch a movie, be on a cell phone on my trip<br><br>scale of 1 (strongly disagree) to 5 (strongly agree) |
|  | 19 | I think Driving in congested areas is stressful<br><br>scale of 1 (strongly disagree) to 5 (strongly agree) |
|  | 20 | I find autonomous vehicles to be useful when I'm not feeling well<br><br>scale of 1 (strongly disagree) to 5 (strongly agree) |
|  | 21 | Using autonomous vehicles will be useful in meeting my driving needs<br><br>scale of 1 (strongly disagree) to 5 (strongly agree) |
|  | 22 | Using an autonomous vehicle would enable me to reach my destination safely<br><br>scale of 1 (strongly disagree) to 5 (strongly agree) |
|  | 23 | I find autonomous vehicles to be useful when I'm impaired<br><br>scale of 1 (strongly disagree) to 5 (strongly agree) |

**List of Tables:**

**List of Figures:**